\documentclass[twocolumn,prl]{revtex4}

\usepackage{graphicx}
\usepackage{rotating}
\usepackage{amsmath}
\usepackage{amsfonts}
\usepackage{amssymb}
\usepackage{enumerate}
\usepackage{longtable}
\usepackage{dcolumn}
\usepackage{bm}

\begin{document}

\newcommand{\be}{\begin{equation}}
\newcommand{\ee}{\end{equation}}
\newcommand{\bn}{\begin{eqnarray}}
\newcommand{\en}{\end{eqnarray}}

\title{Unconventional Mott Transition in K$_{x}$Fe$_{2-y}$Se$_{2}$}

\author{L. Craco$^1$, M.S. Laad$^2$ and S. Leoni$^1$}

\affiliation{$^1$Physical Chemistry, Technical University Dresden,
01062 Dresden, Germany \\
$^2$Institut f\"ur Theoretische Physik, RWTH Aachen University,
52056 Aachen, Germany}

\date{\rm\today}

\begin{abstract}
Whether the newly discovered K$_{x}$Fe$_{2-y}$Se$_{2}$ systems are doped 
Mott or band insulators is key to how superconductivity emerges at lower 
temperature. With extant theoretical studies supporting conflicting 
scenarios, a more realistic approach is urgently called for. Here, 
we use LDA+DMFT to study this issue in K$_{x}$Fe$_{2-y}$Se$_{2}$. We 
find that the undoped KFe$_{1.6}$Se$_{2}$ system is a new kind of Mott-Kondo 
insulator (MKI). Electron doping this MKI drives a Mott transition to 
an orbital-selective non-Fermi liquid metal. Good agreement with 
spectral and transport responses supports our view, implying that 
superconductivity arises from a doped Mott insulator, as in the 
high-$T_{c}$ cuprates.  
\end{abstract}


\maketitle

Recent finding of high-$T_{c}$ supercxonductivity (HTSC) with
$T_{c} \gtrsim 30$~K in K$_{x}$Fe$_{2-y}$Se$_{2}$~\cite{jgou} and 
(K,Tl)Fe$_{x}$Se$_{2}$~\cite{xpwang,fang} is remarkable for the following 
reasons: (i) While its SC $T_{c}$ is comparable to that for the 
1111- and 122-Fe pnictides (FePn), it occurs near antiferromagnet 
(AF) {\it insulators}: Fe composition is the tuning parameter for 
insulator-metal and SC instabilities. Evidence for Fe vacancy 
order~\cite{sab} is interesting: its generic effect is to reduce
LDA (local-density-approximation) bandwidths~\cite{si}. Theoretically, 
both Mott~\cite{si,zhang,dai} and band~\cite{ig} insulating states 
have been proposed as candidates. (ii) It intensifies the fundamental 
debate~\cite{si-nature} on the degree of electronic correlations in 
Fe-based SC. Transport and optical data reveal insulating behavior 
(albeit with small gap) well above $T_{N}(\delta)$, $\delta$ being 
the electron doping. However, optical~\cite{opt} and 
ARPES~\cite{mott-kondo} studies also clearly show large-scale 
spectral weight transfer (SWT) (over an energy scale $O(2.0)$~eV) 
as a function of temperature ($T$) across the magnetic and SC 
instabilities, a fingerprint of Mottness. Both $\rho(T)\simeq T$ 
and bad metallicity above $T_{c}$ are features shared along with 
other non-Landau-Fermi-liquid (LFL) metals close to a Mott or 
selective-Mott instability~\cite{sachdev}. ARPES~\cite{mott-kondo} 
also seems to show a curious co-existence of Mott- and 
``band-insulating'' spectral features in K$_{x}$Fe$_{2-y}$Se$_{2}$. 
Large-scale SWT on a scale of O$(2.0)$~eV in response to a small 
($k_{B}T\simeq$O$(10)$~meV) perturbation, however betrays the 
``hidden'' strong correlations. Its implications for SC are 
intriguing: are the Mott- or ``band''-like subsets of the 
renormalised spectra relevant for SC that emerges at lower $T$? (iii) 
Finding of large local moment value on Fe, ${\bf M}_{Fe}\simeq 3.3\mu_{B}$,
suggests strong electronic correlations. Even in the more {\it itinerant} 
FePn, relevance of a ``dual'' picture for the parent magnets is now
increasingly recognised~\cite{julian}: such a dual picture must be 
even more relevant for K$_{x}$Fe$_{2-y}$Se$_{2}$.
 
These findings {\it necessitate} incorporation of reasonably strong 
multi-band electronic correlations. All five $d$-bands crossing the 
Fermi energy ($E_F$) must be kept at a ``minimally realistic'' level 
in order to satisfactorily resolve the doped Mott-versus-band insulator 
issue above. In multi-orbital (MO) systems, sizable correlations also 
drive new physical effects: they induce orbital selective (OS) 
bad-metallic states with no LFL coherence, naturally yielding 
bad metallic resistivity above $T_{c}$. An OS-Mott scenario also 
generically ``wipes out'' a subset of Fermi surface (FS) sheets from the 
{\it renormalized}, correlated band structure (Lifshitz transition). 
In turn, this can have far-reaching consequences for the symmetry of 
the SC pair function, $\Delta({\bf k})$: details and presence or 
absence of gap nodes crucially depends on whether or not $\Delta({\bf k})$ 
intersects the {\it renormalized} FS of such a metal. Extant ARPES data 
show only electron pockets in metallic K$_{0.8}$Fe$_{1.7}$Se$_{2}$ and 
anisotropic $s$-wave SC gap~\cite{qian}. However, NMR 
$T_{1}^{-1}(T)\simeq T^{2}$ for $T< \frac{1}{2} T_{c}$ remains puzzling in 
this context~\cite{ma}. This unsettled state of affairs calls for 
detailed theoretical scrutiny: it must, in view of the fundamentally 
conflicting views discussed above, base itself on an satisfying 
description of the ``normal'' state.

Here, we use state-of-the-art LDA-plus-dynamical mean-field theory 
(LDA+DMFT)~\cite{kot-rev} to address these issues. In LDA+DMFT studies 
of FePn, their degree of correlatedness has been the bone of contention: 
consensus has fluctuated between weakly~\cite{anisi} to sizably 
correlated limits~\cite{1111-work}. In K$_{x}$Fe$_{2-y}$Se$_{2}$, 
a perusal of the resistivity data~\cite{mott-kondo} for non-AF but 
SC samples reveals that these are proximate to a Mott transition: 
$\rho_{dc}(T)$ immediately above $T_{c}$ is very bad metallic and 
quickly crosses over to an insulator-like dependence. In these cases, 
AF order {\it cannot} be responsible for insulating behavior: rather, 
these data show that destroying AF order reveals underlying Mottness, 
where (i)-(iii) above can be rationalised naturally. Clearly, 
electronic correlations can only get stronger as one approaches 
the insulator, and AF and/or orbital order can naturally arise 
via spin-orbital superexchange. In this work, we confirm this 
hypothesis, showing that a strong correlation view achieves good 
semiquantitative accord with a range of data in {\it both}, 
insulating and metallic phases of K$_{x}$Fe$_{2-y}$Se$_{2}$. In 
particular, we clarify the co-existing Mott- and ``band'' 
insulator-like features in ARPES in a qualitatively new 
Mott-Hubbard scenario. Armed with these strengths, we qualitatively 
discuss the constraints our view imposes on mechanisms of 
(unconventional) SC in K$_{x}$Fe$_{2-y}$Se$_{2}$.

\begin{figure}[t]
\includegraphics[width=3.6in]{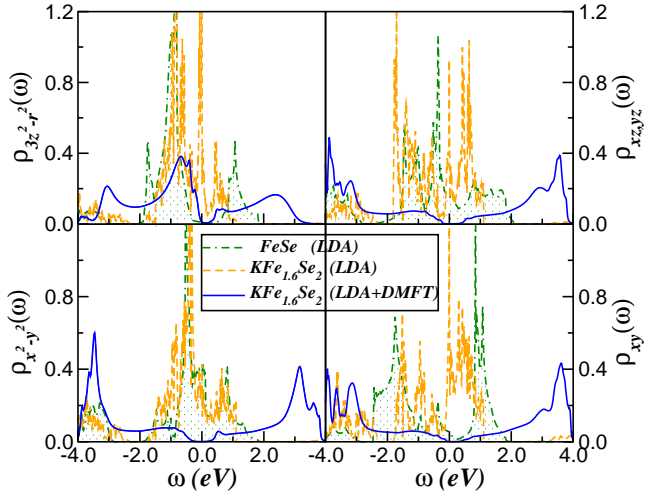}
\caption{
Orbital-resolved LDA  density-of-states (DOS) for the Fe $d$ orbitals of
FeSe (dot-dashed) and KFe$_{1.6}$Se$_2$ (dashed) as well as LDA+DMFT
(with $U=4.0$~eV, $U'=2.6$~eV and $J_{H}=0.7$~eV) for undoped 
KFe$_{1.6}$Se$_2$.}
\label{fig1}
\end{figure}

We start with the experimental structure of KFe$_{1.6}$Se$_2$~\cite{sab}. 
LDA calculations were performed using the linear muffin-tin orbitals 
(LMTO)~\cite{ok} scheme, in the atomic sphere approximation: 
Self-consistency is reached by performing calculations with 242 
irreducible {\bf k}-points. The radii of the atomic spheres were 
chosen as $r$=2.6~(Fe), $r$=4.25~(K) and $r$=2.7~(Se) a.u. in 
order to minimize their overlap. Within LDA,the one-electron part is 
$H_{0}=\sum_{{\bf k},a,\sigma}
\epsilon_{a}({\bf k})c_{{\bf k},a,\sigma}^{\dag}c_{{\bf k},a,\sigma}$, where
$a=xz,yz,xy,3z^{2}-r^{2},x^{2}-y^{2}$ label the (diagonalized in orbital 
basis) five $3d$ bands. The MO Coulomb interactions (treated within DMFT)
constitute the interaction term, which reads
$H_{int}=U\sum_{i,a}n_{ia\uparrow}n_{ia\downarrow}
+ U'\sum_{i,a \ne b}n_{ia}n_{ib} - J_{H}\sum_{i,a,b}{\bf S}_{ia} \cdot 
{\bf S}_{ib}$.  We use the MO iterated perturbation theory as an impurity 
solver for DMFT.  Though not numerically {\it exact}, it has a proven 
record of recovering correct LFL metallic behavior~\cite{raja} and 
good semiquantitative agreement in a host of real systems. 

\begin{figure}[t]
\includegraphics[width=3.6in]{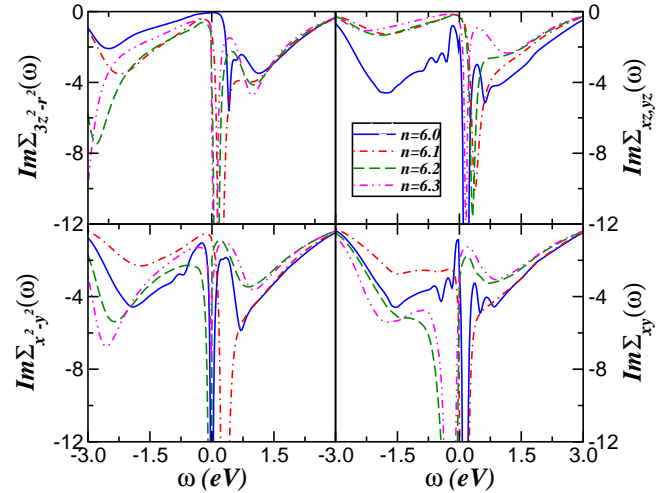}
\caption{
Orbital-resolved (DMFT) imaginary parts of the self-energies for the 
Fe $d$ orbitals of pure and electron doped KFe$_{1.6}$Se$_2$ with 
$U=4.0$~eV, $U'=2.6$~eV and $J_{H}=0.7$~eV. The selective-Mott nature
is clear.}
\label{fig2}
\end{figure}

Fig.~\ref{fig1} shows the LDA DOS for KFe$_{1.6}$Se$_{2}$, whereby a clear
and sizable reduction (O$(20)\%$) of the average LDA bandwidth $(W_{LDA})$ 
relative to that for FeSe, induced by presence of Fe vacancy order, is seen. 
FeSe is already a bad metal close to a Mott insulator~\cite{craco-fese}, and 
the significantly smaller $W_{LDA}$ for KFe$_{1.6}$Se$_{2}$ then naturally 
suggests emergence of a Mott insulator in the latter. Indeed, our results 
show a small but clear insulating gap in DMFT spectra. Several interesting 
features, especially germane to the above discussion, are now manifest: 
(i) the Mott gap is clearly orbital-dependent, i.e, intrinsically anisotropic. 
(ii) Examination of the orbital-resolved (imaginary parts) self-energies 
reveal a behavior hitherto unique to Fe-based systems. Namely, 
Im$\Sigma_{a}(\omega)$ with $a=xz,yz,xy,x^{2}-y^{2}$ clearly reveal 
their Mott insulating character, i.e, a pole at $\omega=E_{F}(=0)$, 
as shown in Fig.~\ref{fig2}. On the other hand, Im$\Sigma_{a}(\omega)$ 
for $a=3z^{2}-r^{2}$ simultaneously shows Kondo insulator features, i.e, 
Im$\Sigma_{a}(\omega)=0$ in the gap region. Thus, remarkably, 
KFe$_{1.6}$Se$_{2}$ shows co-existing Mott and Kondo insulating 
gaps, and we dub this a Mott-Kondo insulator (MKI). One might be tempted to
try to link this to the band insulator state found in LDA, especially 
since a Kondo insulator is an analytically continued version of a band 
insulator (where sizable correlations do exist above the gap scale). 
If electron doping would result in a metal where only the $d_{3z^{2}-r^{2}}$ 
band  would be driven metallic, it would necessarily mandate an
{\it effective}, doped Kondo insulator~\cite{mazin}, rather than doped 
Mott-insulator modelling.  However, in ARPES data for the doped metal, 
the renormalised FS comprises sheets having only $xz,yz$ orbital character, 
in accord with DMFT results (see below). Thus, the insulator-metal 
transition must now be viewed in terms of a doped Mott insulator, in 
accord with earlier model-based work~\cite{si,zhang}. We emphasise 
that no such insulating state is found for $U\leq 3.0$~eV (not shown) 
in absence of magnetic order.   

\begin{figure}[t]
\includegraphics[width=\columnwidth]{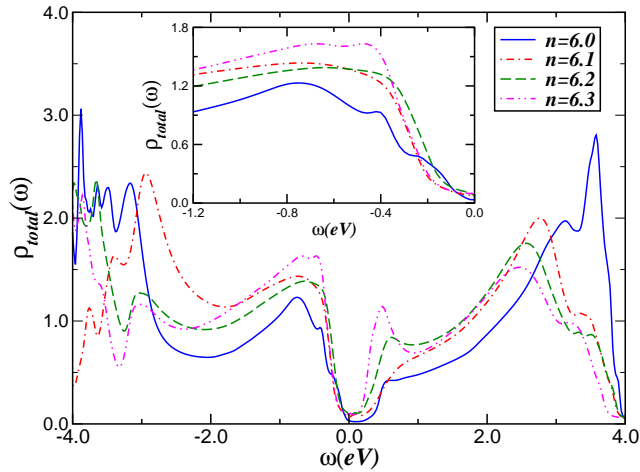}
\caption{
Total DMFT DOS for KFe$_{1.6}$Se$_2$ (main panel) as a function of 
electron doping $n=6+\delta$, clearly showing a low-energy pseudogap 
associated with bad-metallicity. Low-energy DMFT spectra (inset) in 
good agreement with ARPES results~\cite{xpwang,mott-kondo,qian}.}
\label{fig3}
\end{figure}

Electron doping ($n\equiv 6+\delta$, with $\delta>0$) the MKI leads 
to a bad-metallic state. Consistent with our results in
Fig.~\ref{fig2}, we find that only $xz,yz$ orbitals show bad metallic 
behavior, characterised by absence of sharp LFL resonances at $E_{F}$, 
while Mott insulating behavior persists in the $xy,x^{2}-y^{2},3z^{2}-r^{2}$ 
chanels. Thus, we find an OS metal, and strong scattering between the 
Mott-localised and metallic states leads to complete suppression of the 
LFL quasiparticles via the Anderson orthogonality catastrophe~\cite{xre}, 
leading to emergence of anomalously broad spectra in DMFT, see 
Fig.~\ref{fig3}. Microscopically, infrared LFL behavior (narrow 
``Kondo'' resonance in DMFT) is killed off by strong scattering 
between the Mott-localised and quasi-itinerant components of the (DMFT) 
matrix-spectral function, due to sizable $U',J_{H}$, and is intimately 
tied with the OS insulator-metal transition in the five-band Hubbard 
model we use. Thus, our selective-Mott metal is a MO counterpart of 
the FL$^{*}$ metal~\cite{senthil}.

\begin{figure}[t]
\includegraphics[width=\columnwidth]{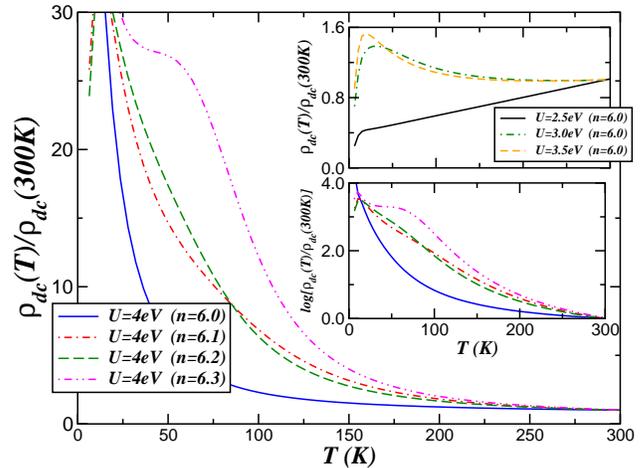}
\caption{
Resistivities for insulating (solid curve) and electron-doped phases of 
K$_{x}$Fe$_{2-y}$Se$_2$ as a function of $T$ in good accord with transport 
data~\cite{fang,res-ref}.}
\label{fig4}
\end{figure}

\begin{figure}[t]
\includegraphics[width=\columnwidth]{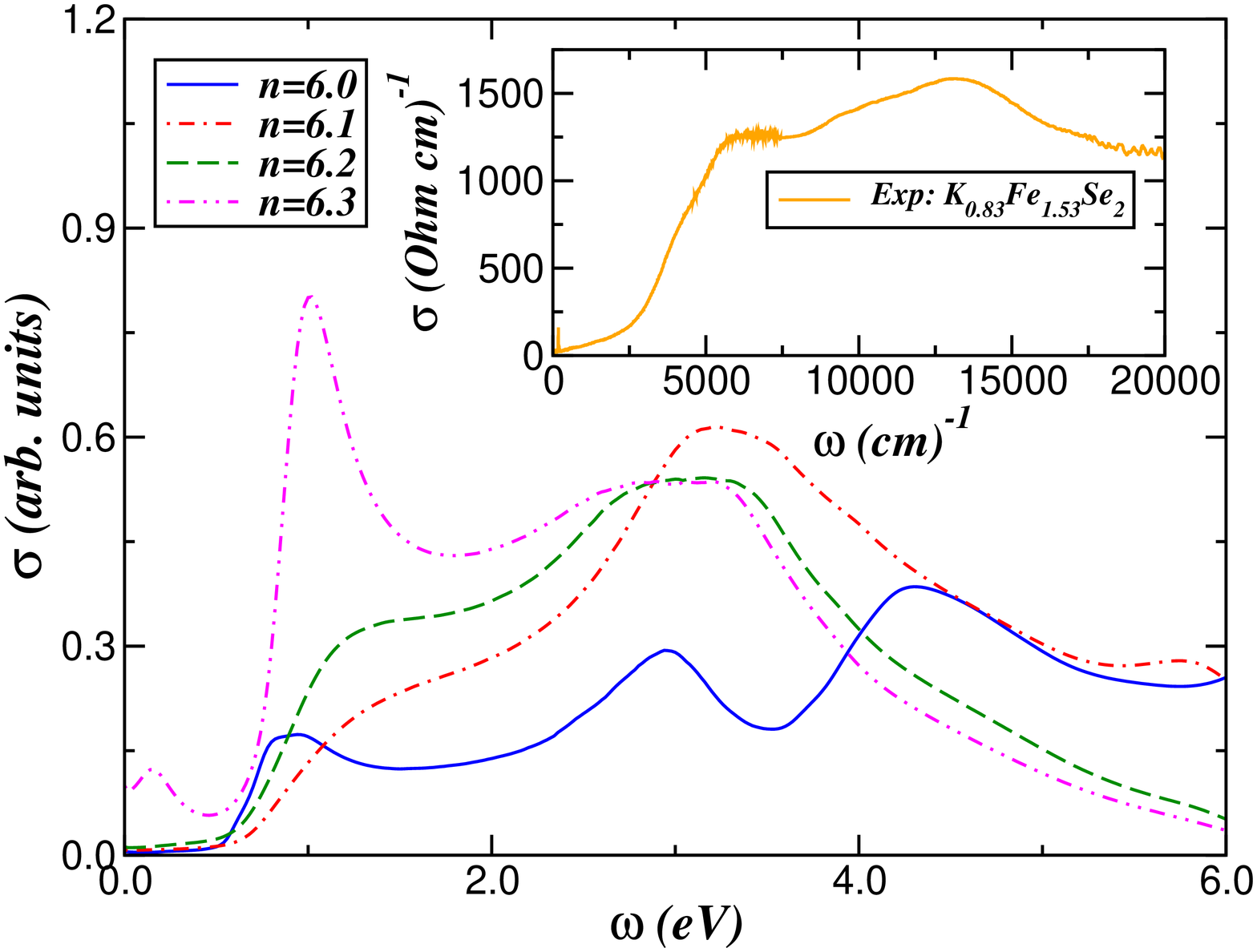}
\caption{(Color online)
Optical conductivity for insulating (solid curve) and electron-doped 
phases of K$_{x}$Fe$_{2-y}$Se$_2$ as a function of energy. The good 
accord with experimental results (inset)~\cite{opt} up to 
$\Omega\simeq O(1.0)$~eV is seen.}
\label{fig5}
\end{figure}

If our proposal is to hold, a range of responses must find a consistent
explication without additional assumptions: we now show this is indeed 
the case. (i) A direct comparison between DMFT spectra and (AR)PES data 
show very good accord: in addition to describing the overall PES lineshape
very well, the DMFT spectra (inset of Fig.~\ref{fig3}) also resolve two 
peaks in the low-energy ($-0.7<\omega<0.0$~eV) range seen in recent 
work~\cite{mott-kondo} for their ``semiconducting'' samples. Further, 
the renormalised FS is composed of predominantly $xz,yz$ orbital states, 
and absence of hole pockets in LDA thus persists in DMFT, in accord 
with ARPES. Absence of LFL quasiparticles naturally implies broad ARPES 
lineshapes without LFL quasiparticles for electron-doped samples, as 
well as sizable SWT over an energy scale $O(1.5-2.0)$~eV in polarised 
X-ray absorption (XAS) studies: this could be tested in future. (ii) 
Transport properties in DMFT are directly computable from the full 
DMFT propagators, since vertex corrections entering the Bethe-Salpeter 
equation are small enough to be neglected in multi-band 
cases~\cite{silke-v2o3}. In Fig.~\ref{fig4}, we show the dc resistivity 
for various $n$. Clear (Mott) insulating behavior obtains for $n=6.0$, 
as expected. With $\delta>0$, a doping-dependent crossover from a 
``high''-$T$ insulator to a low-$T$ (lower inset in Fig.~\ref{fig4}) bad
metal obtains. Obviously, this crossover scale, marked by the maximum 
of $\rho_{dc}(T)$, increases with $\delta$. The upper inset of 
Fig.~\ref{fig4} shows that small $U$ yields metallic behavior up to 
high $T$, in conflict with data. At the very least, $U>3.5$~eV is 
needed to obtain accord for $\delta=0$, but correct insulating behavior 
is only found for $U \cong 4.0$~eV. (iii) Further corroborative support comes 
from optical conductivity. In Fig.~\ref{fig5}, we show the total optical 
conductivity, $\sigma(\omega)$, within DMFT, using the multi-band version 
of the general DMFT formalism~\cite{silke-v2o3}. $\sigma(\omega)$ rises 
with $\omega$ at small $\omega$, and the optical gap, $\Delta\simeq 0.3$~eV, 
is in good accord with data~\cite{opt}. Further, as also seen, 
bad-metallic conductivity, along with large-scale ($>O(2.0)$~eV) transfer 
of dynamical spectral weight, obtains for $\delta>0$. It is particularly 
noteworthy that good accord is seen not only at low energy 
($0<\omega<0.3$~eV), but up to $1.0$~eV, as seen by direct comparison 
with data (inset to Fig.~\ref{fig5}). Comprehensive accord with PES and 
(dc and ac) conductivities thus provides strong support to the view that 
KFe$_{1.6}$Se$_{2}$ is a Mott insulator (MI), and that the doping-driven 
transition to a bad metal must then be viewed in terms of a doped MI.

These findings put strong constraints on mechanisms of SC. Since LFL
quasiparticles are not stable excitations in the bad metal, 
instabilities to ordered states via (particle-particle) BCS-like 
pairing of well defined LFL quasiparticles are untenable. In non-LFL, 
bad metals coherent one-electron hopping term is irrelevant in an RG 
sense~\cite{xre}. Hence, ordered states can only arise via {\it coherent} 
two-particle hopping, which becomes more relevant when one-electron 
hopping term is irrelevant: the situation is analogous to coupled 
$D=1$ Luttinger liquids~\cite{giam}. Given local incoherent metallic 
state(s) within DMFT, the above analogy tells us that residual, 
inter-site and inter-orbital (in multi-band systems) two-particle 
interactions can generate ordered states directly from the bad metal. 
This is the philosophy used earlier~\cite{our-prl} for the 1111-FePn 
systems. As in earlier work, restricting ourselves to the $xz,yz$ 
orbital sector, the effective pair-hopping term (second order in 
$t_{ab}$) is $H^{(2)}\simeq -1/2\sum_{k,k',a,b}V_{ab}(k,k')c_{ka\sigma}^{\dag}c_{b,k',\sigma'}^{\dag}c_{b,k'\sigma'}c_{ka\sigma}$.
Decoupling $H^{(2)}$ in the particle-hole and particle-particle channels gives 
$H_{MF}^{(2)}=\sum_{k,a,b}[(\Delta_{ab}^{(1)}(k)c_{ka\sigma}^{\dag}c_{kb\sigma}+h.c)
+ (\Delta_{ab}^{(2)}(k)c_{ka\uparrow}^{\dag}c_{-kb\downarrow}^{\dag}+h.c)]$, where 
$\Delta_{ab}^{(1)}(k)=\langle \gamma(k)c_{kb\sigma}^{\dag}c_{ka\sigma}\rangle$ and 
$\Delta_{ab}^{(2)}(k)=\langle \gamma(k)c_{kb\sigma}c_{ka-\sigma}\rangle$ with
$\gamma(k)=$cos$k_{x}+$cos$k_{y}+\alpha$cos$k_{x}$cos$k_{y}$ for the frustrated 
case of Fe-based systems.  These represent orbital nematic (with orbital 
order and lattice distortion)~\cite{our-nematic} and inter-orbital 
pairing~\cite{our-prl} instabilities. Extending DMFT to study both 
these orders in K$_{x}$Fe$_{2-y}$Se$_{2}$ is more problematic, however: 
the large moment, $\mu_{Fe}=3.3\mu_{B}$, the block-spin moment 
${\bf M}\simeq 11\mu_{B}$, and the block spin-AF order~\cite{bao} suggests
that both, possible orbital order~\cite{phillips} with lattice distortion 
and subsequent AF, as well as SC instabilities must involve coupling 
between four-site plaquettes, beyond what our $H_{MF}^{(2)}$ would give. 
Since the Mott transition already occurs at high $T$, a way to proceed 
might involve using the present DMFT results as a template for deriving 
an appropriate low-energy, plaquette-centered model using the {\it active} 
$xz,yz$ orbital states to address these issues as recoginsed by Baskaran 
in a different approach~\cite{bask}. This is currently underway, and 
will be reported in future.         

In conclusion, using LDA+DMFT for a minimally realistic five-band Hubbard 
model, we resolve the issue of a doped Mott- vs band insulator physics in
K$_{x}$Fe$_{2-y}$Se$_{2}$ systems in favor of the former, Mott view. 
Good quantitative accord with key spectral and transport data in a sizably 
correlated picture confirms this view, and strongly suggests close 
underlying similarities (in spite of very different chemistry) between SC 
emerging here from a doped multi-orbital Mott-Kondo-insulating state with 
$d$-wave SC in doped high-$T_{c}$ cuprates.

L.C. thanks the Physical Chemistry department at Technical University
Dresden for hospitality.

\end{document}